\documentclass[a4paper,11pt]{article}
\pdfoutput=1 

\usepackage{jheppub} 

\usepackage[T1]{fontenc} 

\title{\boldmath Integral representation of the neutrino mass-squared differences}

 \author{I. Alikhanov}
 \affiliation{North-Caucasus Center for Mathematical Research, North-Caucasus Federal University,\\
Stavropol 355017, Russian Federation}

\emailAdd{ialspbu@gmail.com}

\abstract{Determining the absolute neutrino mass scale remains one of the most compelling challenges in particle physics. To constrain theoretical models, establishing precise relations among neutrino masses is essential. We propose a simple integral representation of the neutrino mass-squared differences $\Delta m_{ij}^2$ that provides a complementary perspective on these oscillation parameters. We then demonstrate its utility through several examples. Specifically, assuming stringent cosmological bounds that confine the sum of neutrino masses near the normal ordering floor, we derive an analytical condition for the lightest neutrino mass,~\(m_1 < \sqrt{61\Delta m^2_{21}}/30\). Using recent data from the JUNO experiment, this yields a competitive upper limit of \(m_1<0.0023~\text{eV}\) (\(95\%\) C.L.). We also formulate practical analytical bounds for~\(m_{2}\) and~\(m_{3}\) adaptable to future data, and translate the  results into allowed ranges for the effective electron and Majorana neutrino masses~$m_{\nu_e}$ and~$m_{\beta\beta}$.  Finally, we show that neutrino mass relations of the Gatto--Sartori--Tonin type emerge directly from the proposed integral representation.}

\begin{document} 
\maketitle
\flushbottom

\section{Introduction}
According to the standard three-neutrino paradigm—assumed throughout this article—each of the three neutrino flavor eigenstates ($\nu_e$,~$\nu_\mu$,~$\nu_\tau$) is a quantum mechanical superposition (mixture) of three neutrino mass eigenstates ($\nu_1$, $\nu_2$, $\nu_3$) with corresponding masses $m_1$, $m_2$, $m_3$. This mixing is described by the unitary $3\times3$ Pontecorvo--Maki--Nakagawa--Sakata matrix, $U_{{l}i}$~\cite{ParticleDataGroup:2024cfk}, which is typically parametrized by three angles~($\theta_{12}$, $\theta_{23}$, $\theta_{13}$) and a CP-violating phase~($\delta_{\text{CP}}$).

An array of experiments with solar~\cite{SNO:2002tuh,Super-Kamiokande:2001ljr}, atmospheric~\cite{Super-Kamiokande:1998kpq,IceCube:2014flw}, accelerator~\cite{K2K:2004iot,MINOS:2008kxu,T2K:2011ypd,OPERA:2015wbl}, and reactor~\cite{KamLAND:2002uet,DoubleChooz:2011ymz,DayaBay:2012fng,RENO:2012mkc} neutrinos has confirmed the existence of oscillations, defined as transitions between the flavor eigenstates, thereby demonstrating that at least two of the three neutrino masses are nonzero. While oscillation experiments have also successfully measured the differences between the squares of the neutrino masses (the solar and atmospheric mass splittings), the absolute mass scale and the ordering of the masses—whether normal ($m_1<m_2<m_3$) or inverted ($m_3<m_1<m_2$)—remain among the most compelling open questions in modern particle physics. However, there is a mild preference for normal ordering~\cite{T2K:2023smv,T2K:2025yoy,NOvA:2025tmb, Esteban:2026phq}. Interestingly, datasets from~\cite{T2K:2023smv,T2K:2025yoy} and~\cite{NOvA:2025tmb}, when combined in the first joint analysis~\cite{T2K:2025wet}, ceased to show a preference for either ordering. The now-operational JUNO detector~\cite{JUNO:2022mxj,JUNO:2025gmd} and upcoming experiments like DUNE~\cite{DUNE:2020lwj} and Hyper-Kamiokande~\cite{Hyper-Kamiokande:2018ofw} are aimed at establishing the actual neutrino mass ordering with high statistical confidence~($3-5\sigma$). There are also complementary, yet model-contingent, cosmological bounds on the sum of neutrino masses~($\Sigma m_{\nu}=m_1+m_2+m_3$) that strongly suggest normal ordering~\cite{DESI:2024mwx, Elbers:2025vlz,AtacamaCosmologyTelescope:2025nti,SPT-3G:2025bzu,Montandon:2026vuc,Jimenez:2026ycn}. For example, the DESI collaboration reports a bound of~$\Sigma  m_\nu<0.072$~eV~(95\%~CL)~\cite{DESI:2024mwx}, which, however, can be relaxed to~$\Sigma m_\nu<0.11$~eV when employing alternative analysis tools or different  dataset combinations~\cite{Naredo-Tuero:2024sgf,Gorbunov:2026sly}. 

Model-independent information on the neutrino mass scale is primarily derived from direct measurement experiments, such as KATRIN~\cite{Drexlin:2026zam}. The latter investigates the endpoint region of the tritium beta spectrum to provide  an upper bound on the effective "electron neutrino mass",~$m_{\nu_e}=\sqrt{\Sigma_i|U_{ei}|^2m^2_i}$. It has recently set the limit of~$m_{\nu_e}<0.45$~eV at~90\%~C.L.~\cite{KATRIN:2024cdt} and is about to reach its target sensitivity of less than~0.3~eV~\cite{Drexlin:2026zam}.  Another relevant observable of fundamental interest is the effective Majorana mass, \mbox{$m_{\beta\beta}=|\Sigma_iU_{ei}^2m_i|$},  probed in searches for neutrinoless double-beta ($0\nu\beta\beta$) decay. In contrast to~$m_{\nu_e}$, the Majorana mass additionally depends upon unknown phases~\cite{DellOro:2016tmg,Gomez-Cadenas:2023vca}. The detection of~$0\nu\beta\beta$ decay would demonstrate that neutrinos are Majorana fermions and  confirm that total lepton number is not conserved.  According to the KamLAND-Zen collaboration,  the upper limits on $m_{\beta\beta}$ are in the range $0.028~\text{eV}-0.122$~eV \cite{KamLAND-Zen:2024eml}, depending on the nuclear matrix element used. This corresponds to a 90\%~C.L. lower limit on the $0\nu\beta\beta$ decay half-life in ${}^{136}\text{Xe}$ of $T^{0\nu}_{1/2}>3.8\times10^{26}$~yr. 
Details of various neutrino measurements are provided, e.g., in~\cite{Nu2024}.

Of the three neutrino mass-squared differences, $\Delta m^2_{21}=m^2_2-m^2_1$, $\Delta m^2_{31}=m^2_3-m^2_1$, and $\Delta m^2_{32}=m^2_3-m^2_2$, only two are conventionally regarded as independent~\cite{Denton:2025jkt}. Notably, it has recently been suggested that an underlying algebraic relationship could reduce the true number of independent parameters even further~\cite{Alikhanov:2026jmw}. On the experimental frontier, the JUNO collaboration has delivered the most precise measurement of \(\Delta m^2_{21}\) to date, reporting a value of \((7.50 \pm 0.12) \times 10^{-5}\)~eV\({}^{2}\)~\cite{JUNO:2025gmd}. Global analyses of neutrino oscillation data continue to place tight constraints on the remaining sector~\cite{Capozzi:2017ipn,Capozzi:2025wyn,Esteban:2024eli,Esteban:2026phq,deSalas:2020pgw}; for instance, the recent NuFIT-6.1 global analysis, which incorporates JUNO’s latest constraints, yields \(\Delta m^2_{31} = (2.529 \pm 0.021) \times 10^{-3}\)~eV\({}^{2}\) for normal ordering\footnote{From now on, we will assume  normal mass ordering.}~\cite{Esteban:2026phq}. These parameters will be measured with sub-percent precision in the foreseeable future, particularly due to JUNO~\cite{JUNO:2022mxj}. In this regard, it becomes important to take into account sub-leading corrections, such as matter effects~\cite{Khan:2019doq}. Crucially, however, these observables and their inherent  consistency condition, $\Delta m^2_{31}-\Delta m^2_{21}=\Delta m^2_{32}$, remain insufficient to determine the absolute neutrino masses, since   the number of unknowns exceeds the number of equations. Therefore, additional  relations among the neutrino mass-squared differences would facilitate  this determination. 
In mathematical analysis a single function can often manifest in multiple, distinct forms---such as Taylor series, integral transforms, Fourier series, or differential equations. Far from mere redundancy, exploring these varied representations is crucial; each framework unlocks unique structural insights, reveals hidden symmetries, and yields practical methods for theoretical proofs and computational problem-solving. While one formulation might easily expose a function's local behavior, another could naturally highlight its global properties or long-term evolution, allowing us to view the same underlying phenomenon from complementary perspectives. Guided by this philosophy, we propose a simple integral representation of the neutrino mass squared differences \(\Delta m^2_{ij}\) and demonstrate its utility through several examples.

\newpage 
\section{Formalism}
 Consider the following simple function:

\begin{equation}
F(\theta)=\frac{1}{m_2-m_1\sin{\theta
}},\label{eq:the function}
\end{equation}
where $m_1$ and $m_2$ are the masses of the neutrino states $\nu_1$ and $\nu_2$, respectively, \mbox{$-\infty<\theta<+\infty$}. Since~$m_2>m_1$, as guaranteed by the measured sign of~$\Delta m^2_{21}$~\cite{JUNO:2025gmd}, there are no values of~$\theta$ for which the denominator in~\eqref{eq:the function} vanishes. Thus,~$F(\theta)$ is a smooth, positive, $2\pi$-periodic function defined on the entire real line.

We note that the solar neutrino mass splitting is equal to the reciprocal of the square of the period average of~$F(\theta)$. Indeed, by means of Cauchy's residue theorem, we find that 
\begin{equation}
\frac{1}{2\pi}\int_0^{2\pi}\frac{d\theta}{m_2-m_1\sin{\theta
}}=\frac{1}{\sqrt{\Delta m^2_{21}}}.\label{eq:the integral}
\end{equation}
Integrals of this type are a common feature in  textbooks on complex variables~(see, e.g.,~\cite{brown2014complex}). Such a representation of~$\Delta m^2_{ij}$ should generally be interpreted as an equation for neutrino masses. While the right-hand side provides an accurately measured oscillation parameter, the left-hand side is a well-studied mathematical object—the Riemann integral—whose rich mathematical  properties can be leveraged to evaluate neutrino masses. Several examples using~\eqref{eq:the integral} are given below.

\section{Neutrino mass bounds}
To give an example of the application of~\eqref{eq:the integral}, we proceed as follows. We employ the standard trapezoidal rule and approximate the integral with 12 uniform subintervals (though other partitions may be explored). The step size~$\pi/6$ defines the grid points~$\theta_k=k\pi/6$ for $k=0,\ldots,12$,  resulting in

\begin{equation}
\frac{m_2}{6}\left(\frac{1}{\Delta m_{21}^2}+\frac{8}{m_2^2+3\Delta m_{21}^2}+\frac{8}{3m_2^2+\Delta m_{21}^2}+\frac{1}{m_{2}^2}\right)+r=\frac{1}{\sqrt{\Delta m^2_{21}}},\label{eq:the integral_approx}
\end{equation}
where $r$ denotes the error of the approximation. 
Equation~\eqref{eq:the integral_approx} is an alternative, algebraic form of~\eqref{eq:the integral}. This allows us to directly relate $m_2$ to~$\sqrt{\Delta m^2_{21}}$ and constrain the corresponding proportionality constant.  

In estimating the error $r$, one should be careful. The trapezoidal rule is typically an excellent choice for the numerical integration of a smooth periodic function over a full period, as the error decays exponentially. However, our integrand generally features a peak around \(\theta=\pi/2\), which can disrupt this exponential convergence under certain conditions. The peak provides the contribution to the approximation scaling as $\sim m_2/\Delta m^2_{21}$.  The relative error is straightforwardly estimated to be 
\begin{equation}
|r|\sqrt{\Delta m^2_{21}}<\frac{1}{32}\left(1-\frac{m_2}{\sqrt{\Delta m^2_{21}}}\right)^6.
\label{error bound}
\end{equation}
The error decreases sharply as $m_2$ approaches $\sqrt{\Delta m^2_{21}}$. For instance, the left-hand side of~\eqref{error bound}  remains  below~$3.8\times10^{-11}$ at 95\% C.L., assuming the cosmological total neutrino mass limit of~$\Sigma m_i< 0.0642$~eV~\cite{Elbers:2025vlz,Jimenez:2026ycn}. Thus, the integral representation~\eqref{eq:the integral}  yields a rapidly convergent relation in this region, directly linking $m_2$ to $\sqrt{\Delta m^2_{21}}$  without the explicit use of the lightest neutrino mass.
 To ensure a conservative treatment, we set $r\sqrt{\Delta m^2_{21}}=~-3.8\times10^{-11}$ and thereby intentionally overestimate the magnitude of the  error term. This approach enhances the robustness of our analysis and simultaneously turns~\eqref{eq:the integral_approx} into the strict inequality
\begin{equation}
\frac{m_2}{6}\left(\frac{1}{\Delta m_{21}^2}+\frac{8}{m_2^2+3\Delta m_{21}^2}+\frac{8}{3m_2^2+\Delta m_{21}^2}+\frac{1}{m_{2}^2}\right)<\frac{1+3.8\times10^{-11}}{\sqrt{\Delta m^2_{21}}}.\label{eq:inequality}
\end{equation}
A tight constraint on the mass of the neutrino state $\nu_2$ immediately follows from~\eqref{eq:inequality}:
\begin{equation}
\hskip 0.8cm\sqrt{\Delta m^2_{21}}\leq m_2<\frac{31}{30}\sqrt{\Delta m^2_{21}}.\label{eq:bound}
\end{equation}
From~\eqref{eq:bound} and the definitions of $\Delta m^2_{21}$ and $\Delta m^2_{31}$, we readily obtain upper bounds on the lightest neutrino mass ($m_1$), as well as on the heaviest one  ($m_3$), in an equally accessible and practical algebraic form:
\begin{equation}
\hskip 3.cm m_1<\frac{1}{30}\sqrt{61\Delta m^2_{21}},\label{eq:bound_light}
\end{equation}
\begin{equation}
\hskip 2.15cm\sqrt{\Delta m^2_{31}}\leq m_3<\sqrt{\Delta m^2_{31}+\frac{61\Delta m^2_{21}}{30^2}}.\label{eq:bound3}
\end{equation} 
Using the recent JUNO measurement of $\Delta m^2_{21}$~\cite{JUNO:2025gmd} alongside the NuFIT-6.1 global analysis data for~$\Delta m^2_{31}$~\cite{Esteban:2026phq}, these bounds yield at~95\%~C.L.:
 \begin{equation}
\hskip 2.15cm m_1<0.0023~\text{eV},\label{eq:bound1}
\end{equation}
\begin{equation}
0.0085~\text{eV}\leq m_2<0.0091~\text{eV},\label{eq:bound2}
\end{equation}
\begin{equation}
0.0497~\text{eV}\leq m_3<0.0508~\text{eV}.\label{eq:bound33}
\end{equation}
Accordingly, the total mass is also constrained to
\begin{equation}
\hskip 2.25cm 0.0582~\text{eV}\leq \Sigma m_{\nu}<0.0622~\text{eV}\,\,\,\,\,\,\,\,(95\%~\text{C.L.}).\label{eq:bound4}
\end{equation}
The respective 95\% C.L. ranges for the effective Majorana and electron neutrino masses are found to be
\begin{equation}
\hskip 1.4cm 0\leq m_{\beta\beta}<0.0057~\text{eV},\label{eq:bound44}
\end{equation}
\begin{equation}
0.0085~\text{eV}< m_{\nu_e}<0.0096~\text{eV}.\label{eq:bound55}
\end{equation}
In evaluating these ranges,  the neutrino mixing angles from the NuFIT~6.1 dataset~\cite{Esteban:2026phq} were employed.

The obtained results are competitive, providing either an improvement upon or further validation of the limits recently established in the literature (cf., e.g., \cite{Elbers:2025vlz,Jimenez:2026ycn}). Notably, our analytical conditions are also in good agreement with the predictions of~\cite{Fritzsch:2006sm}, where the neutrino masses were calculated within an alternative approach assuming a parallelism between quark and lepton flavor mixing. The authors obtained $m_1\simeq\sqrt{0.2\Delta m^2_{21}}$, \mbox{$m_2\simeq\sqrt{1.2\Delta m^2_{21}}$}, and $m_3\simeq\sqrt{\Delta m^2_{31}+0.2\Delta m^2_{21}}$. In particular, the mass $m_3$ turns out to be largely insensitive to the value of $m_1$, meaning that the uncertainties in our relation~\eqref{eq:bound3} are primarily driven by the current experimental errors on $\Delta m^2_{31}$. Anticipated high-precision data on the oscillation parameters, especially from experiments  like JUNO~\cite{JUNO:2022mxj}, will tighten these bounds, bringing us closer to pinning down the absolute mass scale of neutrinos.

While our discussion focused specifically on the parameter $\Delta m^2_{21}$ in~\eqref{eq:the integral} as a primary example, the mathematical approach is general and applies equally to $\Delta m^2_{31}$. For this case, it is enough to  replace the index~2 by~3 in~\eqref{eq:the integral} (under the assumption of normal ordering). 
The trapezoidal rule allows the integral approximation error, $r$, to be made arbitrarily small by increasing the number of partitions. This yields  an expression similar to~\eqref{eq:the integral_approx}:~$ m_3(\ldots)/n+r=1/\sqrt{\Delta m^2_{31}}$, where $n$ is a positive integer determined by the partition. By the way, the rule becomes considerably more accurate in this case because $\sqrt{\Delta m^2_{31}}$ is nearly six times larger than $\sqrt{\Delta m^2_{21}}$, resulting in a broader peak at $\theta=\pi/2$. In the limit of vanishing error, we obtain an equation whose only  solution is $m_3=\sqrt{\Delta m^2_{31}}$, which implies $m_1=0$. Our approach thus highlights the configuration with a massless neutrino state~$\nu_1$ as a viable candidate for the physical mass spectrum. This property aligns with~\eqref{eq:bound3}, as well as with independent theoretical considerations~\cite{Alikhanov:2026jmw} and the tendency in recent cosmological studies pushing the sum of neutrino masses, $\Sigma m_\nu$, toward the normal ordering floor~\cite{DESI:2024mwx,Montandon:2026vuc,Jimenez:2026ycn}. At this point, it is worth noting that a solution where~\mbox{$m_1 \ll \sqrt{\Delta m^2_{21}}$} is, in practice, indistinguishable from one where $m_1 = 0$.~\cite{Formaggio:2021nfz}.

It should also be emphasized that the function~$F(x)$ in~\eqref{eq:the function} is not the unique choice. Alternative integrands yielding different algebraic combinations of the neutrino oscillation parameters exist. Employing such functions and applying more sophisticated methods to approximate the definite integral may lead to more stringent bounds on neutrino masses. 
\section{Neutrino mass relations}
As a further application of~\eqref{eq:the integral}, we  outline how to rigorously and rapidly establish various  neutrino mass relations of the Gatto--Sartori--Tonin type~\cite{Gatto:1968ss}. Such relations are conventionally introduced for neutrinos by assuming analogies with the quark sector \cite{Fritzsch:2006sm}. Meanwhile, they also emerge directly from the proposed integral representation itself. To see this, it is sufficient to apply the first mean value theorem for integrals to~\eqref{eq:the integral}. The integrand~$F(\theta)$   evidently satisfies the conditions of the theorem. Therefore, the theorem  guarantees the existence of an angle~$\theta'$ in the interval $(0, 2\pi)$ such that $\sin\theta'=m_1/(m_2+\sqrt{\Delta m^2_{21}})$ independent of any model assumptions. Utilizing alternative integrals that also yield the parameters $\Delta m^2_{21}$ or $\Delta m^2_{31}$ and applying  mean value theorems lead to other relations between neutrino masses and fixed angles. Remarkably, determining  such a specific angle immediately resolves the  neutrino mass spectrum. In other words, the challenge of finding neutrino masses reduces entirely to evaluating this single geometric parameter. This problem will be investigated in more detail elsewhere.

\section{Conclusions}
Determining the absolute neutrino mass scale remains one of the most compelling challenges in particle physics. To constrain theoretical models, establishing precise relations among neutrino masses is essential. We have proposed a simple integral representation of the neutrino mass-squared differences~$\Delta m^2_{21}$ and~$\Delta m^2_{31}$ that provides a complementary perspective on these oscillation parameters. Applying this representation and assuming stringent cosmological bounds that confine the sum of neutrino masses near the normal ordering floor, we have derived an analytical condition for the lightest neutrino mass,~$m_1 < \sqrt{61\Delta m^2_{21}}/30$. Using recent data from the JUNO experiment, this yields a competitive upper limit of~$m_1<0.0023~\text{eV}$~(95\% C.L.). We have also formulated practical analytical bounds for~$m_{2}$ and~$m_{3}$ adaptable to future data, and translated the  results into allowed ranges for the effective electron and Majorana neutrino masses~$m_{\nu_e}$ and~$m_{\beta\beta}$.  Notably, our approach identifies the configuration with a massless neutrino state~$\nu_1$ as a viable candidate for the physical neutrino mass spectrum.
Finally, we have shown that neutrino mass relations of the Gatto--Sartori--Tonin type emerge directly from the proposed integral representation. While normal ordering was assumed throughout, the extension  to the inverted ordering scenario is straightforward and algorithmic.

The extensive mathematical properties and theorems of the Riemann integral make this framework effective for analyzing neutrino masses and uncovering novel relations among them. Consequently, the integral representation of the oscillation parameters appears promising and warrants detailed theoretical investigation.


\acknowledgments
The author has benefited from correspondence with J.~F.~Beacom.


\bibliographystyle{JHEP}
\bibliography{refs}
\end{document}